\documentclass[a4paper,12pt]{article}
\usepackage{epsfig}
\usepackage{amssymb}
\begin{document}
\thispagestyle{empty}

\newcommand{\etal}  {{\it{et al.}}}  
\def\Journal#1#2#3#4{{#1} {\bf #2}, #3 (#4)}
\def\PRD{Phys.\ Rev.\ D}
\def\NIMA{Nucl.\ Instrum.\ Methods A}
\def\PRL{Phys.\ Rev.\ Lett.\ }
\def\PLB{Phys.\ Lett.\ B}
\def\EPJ{Eur.\ Phys.\ J}
\def\IEEETNS{IEEE Trans.\ Nucl.\ Sci.\ }
\def\CPCD{Comput.\ Phys.\ Commun.\ }

\bigskip

{\bf
\begin{center}
 \textbf{\large {Correlations of two photons at hadron colliders} }
\end{center}
}


\begin{center}
{\large G.A. Kozlov  }
\end{center}


\begin{center}
\noindent
 {
 Bogolyubov Laboratory of Theoretical Physics\\
 Joint Institute for Nuclear Research,\\
 Joliot-Curie st., 6 Dubna\\
 141980 Moscow region, Russia
 }
\end{center}

\begin {abstract}
\noindent
{
We study the Bose-Einstein  correlations of two photons and their coherent properties  that
can provide the information about the space-time structure of the emitting source through the Higgs-boson decays into two photons. We argue that such an investigation could probe the Higgs-boson mass. The model is rather sensitive to the temperature of the environment and to the external distortion effect in medium.}
\end {abstract}



\begin{center}
\noindent
\end{center}


 {\it Introduction.-} The hadron colliders Tevatron and the LHC are at the stage to provide particle physicists with treasure of data.
These data allowed a precise measurement of many important parameters of  particle physics
in order to test their consistency,
and to discover new particles, in particular, the Higgs-boson. Today little is known about the interplay between the Higgs and gauge bosons in medium. The method of Bose-Einstein (BE) correlations is clear and undeniable part of this physics, complicating the quantum statistical description of multiparticle final states (for review of BE correlations see [1]).

Historically, the BE correlations measurements have concentrated on pion pairs correlations, however have also been applied to more heavy hadrons and gauge bosons.


Correlations of two photons can provide information about the space-time structure of  hot
 matter prior to freeze-out. However, the BE correlations with direct (primary) photons is faced to difficulties
 compared to hadrons primarily due to the small yield of photons emitted directly from the hot region  just after particle collisions. The main background is related with decays of  hadrons into photons.
In order to reject most of the hadron background, the reconstructed events in a spectrometer were required to have a deposit energy (level) of greater than the energy well above the minimal energy at which photons emerge from the hadrons.
Experimentally, the $\pi^{0}$ background constitutes a major difficulty which, however, to some extent,
can be taken care of by measuring the photon pair invariant mass.

In Higgs searching experimental program, it is supposed that two photons are produced mainly through the decay of light Higgs-boson $H$. The $H\rightarrow \gamma\gamma$ decay mode is one of the promising discovery channels for the Standard Model (SM) Higgs-boson in low mass region ($114 < M_{H} < 150$ GeV). Despite the small branching ratio ($\sim 10^{-3}$ for $M_{H} = 120$ GeV) this channel has a simple signature and a very good mass resolution ($\simeq 1.5$ GeV).
From the theoretical point of view the decay of the Higgs into two photons emerges through the one-loop  composed of either fermions (e.g., quarks) or vector bosons, or the  scalar particles (see, e.g.,  [2] and the references therein).

The two-photon BE correlations in Higgs physics can provide a powerful tool to estimate the Higgs-boson mass via the characteristic stochastic scale $L_{st}$, related to correlation radius, which defines the geometrical size of the two-photon  source. The relative correlation function is strongly dependent on the space-time properties of the Higgs decays.

We suppose the gamma-quanta are created by some random currents (sources)
which exist in a restricted space-time region and they are chaotically and randomly disturbed by external fields (forces). The restricted source area is characterized by the scale $L_{st}$, the meaning of which is explained in [3].

In the quark-loop scheme, the non-relativistic bremsstrahlung formula for the current of one photon emission (with four-momentum $p = (p^{0},\,\vec {p})$) is
$$j^{\lambda}(p) =\frac{i\,g}{m_{q}\,p^{0}}\,\vec{q}\cdot\vec{\epsilon}^{\lambda}(p)\,e^{-p^{0}/\varepsilon_{0}}, $$
where $\vec{q}$ is the difference between the spatial momenta of a quark and an antiquark with the mass $m_{q}$, $\vec{\epsilon}_{\lambda}(p)$ is the vector of linear polarization of a photon; $g$ is the coupling constant; $\varepsilon_{0}$ means the phenomenological parameter which depends on the initial energy.
For two sources the transition current is
$$ J^{\lambda}(p) =\sum_{n=1}^{2} e^{i\,p\,y_{n}}\,j^{\lambda}_{n} (p).$$
Index $n$ labels the independent quark-antiquark interactions taking place at different space-time points $y_{n}$ which are considered to be randomly distributed in the region of the (photon) source.

By studying BE correlations of identical particles, it is possible
to determine the time scale and spatial region over which particles do not
have the interactions. Such a surface is called as decoupling one.
In fact, for an evolving system, such as one for  $p p$ collisions, it is
not really a surface, since at each time there is a spread out
surface due to fluctuations in the last interactions, and the shape
of this surface evolves even in time. 

Let us consider the distribution function
 \begin{equation}
\label{e1}
 N_{12}(p_{1}, p_{2}) =
\langle\vert J^{\lambda_{1}}(p_{1})\,J^{\lambda_{2}}(p_{2})\vert^{2}\rangle
\end{equation}
normalized to the product $N_{1}(p_{1})\cdot N_{2}(p_{2})$ with
one-particle distribution functions $N_{i}(p_{i}) =
\langle\vert J^{\lambda_{i}}(p_{i})\vert^{2}\rangle$ ($i=1,2$).
The function (\ref{e1}) defines the probability to find two photons with momenta $p_{1}$ and $p_{2}$ issued at the points $y_{1}$ and $y_{2}$. The crossing momenta has to be taken into account.


In this paper we suggest the  proposal for the  measurements of quantum correlations between two photons at hadron colliders.
This is theoretically supported by the quantum field theory model approaches at finite temperature [4,3],  where one of the main parameters is the temperature of the particle emitting source.

We imply that photons do not strongly interact with the medium: they carry
information about early stage of reaction. Our pragmatic definition is: photons are
produced not from hadronic decays.
The gamma-quantum can be measured from the virtual yield which is observed as some mass of the spectrum for lepton-antilepton pair. The main channels are:
$pp\rightarrow H \rightarrow\gamma^{\star}\gamma^{\star}\rightarrow
2\mu^{-}2\mu^{+},~2e^{-}2e^{+},~e^{-}e^{+}\mu^{-}\mu^{+},~...$ in, e.g., $pp$ collisions.



 {\it BE correlation function. -} In general, a pair of identical bosons with momenta $p_{1}$ and $p_{2}$ and the
mass $m$ produced incoherently
from an extended source will
have an enhanced probability $C_{2}(p_{1},p_{2})=
N_{12}(p_{1},p_{2})/[N_{1}(p_{1})\cdot N_{2}(p_{2})]$ to be measured
in terms of differential cross section $\sigma$, where
 $$N_{12}(p_{1},p_{2})=\frac{1}{\sigma}\frac{d^{2}\sigma}{d\Omega_{1}\,d\Omega_{2}} $$
to be found close in 4-momentum space $\Re_{4}$ when detected
simultaneously, as compared to if they are detected separately with
$$ N_{i}(p_{i})=\frac{1}{\sigma}\frac{d\sigma}{d\Omega_{i}}, \,\,\,
 d\Omega_{i}=\frac{d^{3}\vec p_{i}}{(2\pi)^{3}\,2E_{p_{i}}}, \,\,
 E_{p_{i}}=\sqrt {\vec p_{i}^{2}+m^{2}},\,\,\,
 i = 1, 2. $$
In an experiment, one can account the inclusive density $\rho_{2}(p_{1}, p_{2})$
which describes the distribution of two particles in $\Omega$ (the sub-volume of the phase space)
irrespective of the presence of any other particles
$$ \rho_{2}(p_{1}, p_{2}) = \frac{1}{2!}\frac{1}{n_{events}}\,\frac{d^{2} n_{2}}{dp_{1}\,dp_{2}},$$
where $n_{2}$ is the number of particles counted in a phase space region
$(p_{1} + dp_{1},p_{2} + dp_{2})$. The multiplicity $N$ normalizations stand as
$$\int_{\Omega} \rho(p)\,dp = \langle N\rangle, $$
$$\int_{\Omega} \rho_{2}(p_{1}, p_{2})\,dp_{1}\,dp_{2} = \langle N (N-\delta_{12})\rangle, $$
where $\langle N\rangle$ is the averaged number of produced particles (mean multiplicity). Here, $\delta_{12} =0$
for different particles, while $\delta_{12} =1$ in case of identical particles (coming from the same event).

The following relation can be used to retrieve the BE correlation function $C_2(Q)$:
\begin{equation}
\label{e23}
 C_2(Q) = \frac{N(Q)}{N^{ref}(Q)},
\end{equation}
where $N(Q)$, in general case, is the number of particle pairs
(off-shell photons) in BE correlation pattern with
\begin{equation}
\label{e24}
 Q = \sqrt {-(p_1-p_2)_{\mu}\cdot (p_1-p_2)^{\mu}}= \sqrt{M^{2} - 4\,m^{2}}.
\end{equation}
In definitions  (\ref{e23}) and  (\ref{e24}),  $N^{ref}$ is the number of particle pairs without BE correlations and
$M = \sqrt {(p_1+p_2)^{2}_{\mu}}$ is the invariant mass of the pair of photons.

An essential problem is the estimation of the
reference distribution $N^{ref}(Q)$ in  (\ref{e23}). If there are other
correlations besides the BE ones,  $N^{ref}(Q)$ should be
replaced by a reference sample corresponding to two-particle distribution
in a geometry without BE correlations. Hence, the expression (\ref{e23}) represents the ratio
between the number of $\gamma^{\star}\gamma^{\star}$ pairs $N(Q)$ in the real world and the reference sample
$N^{ref}(Q)$ in the imaginary world. Note, that $N^{ref}$  can not be directly
observed in an experiment. Different methods are usually applied for the construction
of the reference samples [1], however all of them have strong restrictions. One of the preferable
methods
is to construct the reference samples directly from data.
The $\gamma^{\star}\gamma^{\star}$ correlations can be estimated for each bin of the photon average
transverse momentum $p_{T} = \vert \vec {p}_{T_{1}} + \vec {p}_{T_{2}}\vert /2$ as the ratio
of the distribution of photon pair invariant relative momenta where both photons with
transverse momenta $\vec {p}_{T_{1}}$ and  $\vec {p}_{T_{2}}$ were taken from the same event to
the same distribution but with the photons of the pairs taken from different (mixed) events.

In general, the shape of $C_2(Q)$ is model dependent.
The most simple form for $C_2(Q)$ is often used for experimental data fitting [5]
\begin{equation}
 \label{e25}
 C_2(Q)=C_0\cdot [1+\lambda \,e^ {-(Q\cdot R)^{\kappa}}]\cdot (1+\varepsilon Q) ,
\end{equation}
where $C_0$ is the normalization factor, $\lambda$ is the
coherent strength factor, meaning $\lambda =1$ for fully
incoherent and $\lambda =0$ for fully coherent sources; the
symbol $R$ is often called as the "correlation radius", associated with the measure of the particle source size.   The linear term in (\ref{e25}) is supposed to be account within the long-range correlations outside the region of BEC. For the experimental data fitting it is often used either $\kappa =1$ or  $\kappa =2$.
It is  assumed that the maximum of $C_2(Q)$ is 2 for $\vec p_{1} = \vec p_{2}$ if no any distortion and final state interactions are taking into account.
Note that the distribution of bosons
can be either far from isotropic, usually concentrated in some directions, or
almost isotropic, and what is important that in both cases the particles
are under the random chaotic interactions caused by other fields in thermal
medium. In the parameterization (\ref{e25}) all of these issues
should be embedded in $\lambda$.

However, it is known that measured both the size of the source and $\lambda$ are dependent on the transverse momentum $p_{T}$ (or the transverse mass $m_{T}$) and the multiplicity $N$ [6]. The model which predicts the $p_{T}$ ($m_{T}$)-, $N$-  and the temperature  $T$ - dependencies as well as the stochastic forces influence is the so-called Stochastic Evolution  Model at finite temperature, $SEM_{\beta}$, [3]. In this article we develop this model further and apply it to reconstruction of the two-photon correlations. The two-particle correlation function in $SEM_{\beta}$ is (the correlations at large $Q$ are neglected)
 \begin{equation}
 \label{e26}
C_2(p_{1},p_{2}) \simeq \xi(N) \left \{1 + \lambda_{1}(\beta)\,e^{-\Delta_{p\Re}}
\left [1+\lambda_{2}(\beta)\,e^{ \Delta_{p\Re}/2}\right ]\right\} ,
\end{equation}
where $\Delta_{p\Re} =  -(p_{1} - p_{2})^{\mu}\,\Re_{\mu\nu}\,(p_{1} - p_{2})^{\nu}$ is the smearing smooth  generalized function. $\Re_{\mu\nu}$ is the nonlocal structure tensor of the space-time size and it defines the region of emitted photons.
$\xi(N)$ depends on $N$ as $ \xi(N)= \langle{N (N-1)}\rangle/\langle N\rangle^2$.
The function $ \lambda_{1} (\beta)$  is  the measure of the strength of BE correlation between two photons, $\lambda_{1} (\beta)\simeq \gamma(\omega,\beta)/(1+\alpha)^{2}$, and the correction 
is $\lambda_{2}(\beta) \simeq 2\,\alpha/\sqrt{\gamma(\omega,\beta)}$.
 The function $\gamma (\omega,\beta)$ calls for  quantum thermal features of BE correlation
pattern and is defined as
\begin{equation}
\label{e27}
 \gamma (\omega,\beta) \equiv \gamma (n)  = \frac{{n^2 (\bar \omega )}}{{n(\omega )\ n(\omega
 ')}} ,\ \
 n(\omega ) \equiv  n(\omega ,\beta ) =
 \frac{1}{{e^{\omega \beta} - 1 }} ,\ \
 \bar\omega  = \frac{{\omega  + \omega '}}{2} ,
\end{equation}
where $n(\omega,\beta )$ is the mean value of quantum numbers for Bose-Einstein
statistics particles with the energy $\omega$ in the thermal bath
with statistical equilibrium at the temperature $T= 1/\beta$. The
following condition $\sum_{j} n_{j}(\omega,\beta) = N$ is evident,
where the discrete index $j$ stands for the one-particle state.

The function $\alpha \equiv\alpha (\beta, m)$ entering $\lambda_{1}$ and $\lambda_{2}$ in (\ref{e26}), the measure of stochastic forces,
summarizes our knowledge of other than space-time characteristics of the particle
emitting source, and it varies from $0$ to $\infty$.

In terms of time-like $L_{0}$, longitudinal $L_{l}$ and transverse $L_{T}$ components of the
space-time size $\sqrt {L^{2}_{\mu}}$, the distribution $\Delta_{p\Re}$ in (\ref{e26})  looks like
\begin{equation}
\label{e266}
\Delta_{p\Re}\rightarrow \Delta_{pL} = (\Delta p^{0})^{2}\,L_{0}^{2} +
(\Delta p^{l})^{2}\,L_{l}^{2} + (\Delta p^{T})^{2}\,L_{T}^{2},
\end{equation}
where $L_{0}$  is treated as the measure of the particle emission time, or even it represents the interaction strength of outgoing particles.

 Hence, we have introduced a new parameter $\sqrt {L^{2}_{\mu}}$, which defines the region
 of nonvanishing particle density with the space-time extension of the particle emission source.
 Formula (\ref{e26}) must be understood in the sense that $ \exp (-\Delta_{p\Re})$ is a distribution that  in the limit $L\rightarrow\infty $ strictly becomes a $\delta$ - function. For practical using with  ignoring the energy-momentum dependence of $\alpha$, one has
\begin{equation}
\label{e26666}
C_2(Q) \simeq \xi(N) \left  [1 + \frac{\gamma (n)}{(1+\alpha)^{2}} \,e^{-Q^2 L_{st}^{2}} +
\frac{2\,\sqrt{\gamma (n)}\,\alpha}{(1+\alpha)^{2}} \,e^{-Q^2 L_{st}^{2}/2}\right ],
\end{equation}
where the stochastic scale $L_{st}\equiv L_{st} (\beta)$  is the measure of the space-time overlap between two photons, and the physical meaning of $L_{st}$ depends on the fitting of $C_{2}(Q)$-function.
$L_{st}$ can be defined through the evaluation of the root-mean-squared momentum $Q_{rms}$ as:
$$Q_{rms}^{2} (\beta) =\langle \vec Q^{2}\rangle =
\frac{\int_{0}^{\infty} d\vert\vec Q\vert\, \vec Q^{2}\,\left [\tilde C_{2}(Q,\beta) -1 \right ]}
{\int_{0}^{\infty} d\vert\vec Q\vert\,\left [\tilde C_{2}(Q,\beta) -1 \right ]},\,\,\,\,
\tilde C_{2}(Q,\beta) =\frac{C_{2}(Q,\beta)}{\xi (N)},$$
where $L_{st}$ and $Q_{rms}(\beta)$ are related to each other by means of
$$L_{st}(\beta) = \left [\frac{3}{2}\left (1+\frac{1}{1+\frac{1}{4\,\alpha(\beta)}\,\sqrt \frac{{\gamma (n)}}{2}}\right )
\right ]^{1/2}\frac{1}{Q_{rms}(\beta)}. $$
The following restriction $\sqrt {3/2} < (L_{st}\cdot Q_{rms}(\beta)) < \sqrt{3}$ is evident, where
the lower bound satisfies to the case $\alpha\rightarrow 0$ (no any distortion in the particle source), while the upper limit is given by the very strong influence of chaotic forces,
$\alpha\rightarrow \infty$. The result is rather stable in the wide range of variation of $\alpha$.

{\it In-medium distortion.-} It has been emphasized [3] that there are two different scale parameters when the BE correlations are studied. One of them is the correlation radius $R$ introduced in (\ref{e25}). In fact, the latter
gives the "pure" size of the particle emission source.
The other  parameter is the scale $L_{st}$ of the production particle region where the stochastic, chaotic distortion due to environment (medium) is enforced.
$L_{st}$ carries  the dependence of the particle mass, the $\alpha$-coherence degree and  the
temperature.

The question arises: how can BE correlations be used to determine the effective stochastic scale $L_{st}$ and, perhaps, the phase transition?
We suppose the changes of $\gamma^{\star} \gamma^{\star}$ production region and effects
yielding the dynamical variables and parameters of BE correlations due to in-medium distortion.

 In the Higgs-boson rest frame, there is a kinematical configuration
for two pairs of final lepton, antilepton momenta $(p_{l}, p_{\bar l}^{\prime})$ and
$(q_{l}, q_{\bar l}^{\prime})$: $p_{l}\simeq p_{\bar l}^{\prime}$ and
$q_{l}\simeq q_{\bar l}^{\prime}$. The final state mimics a two-body final state,
and if the leptons being the electrons or  muons, the virtual photons with momenta $p$ and
$p^{\prime}$ are nearly on mass-shell, $p^{2}\simeq p^{{\prime}^{2}} \simeq 4\, m^{2}_{l}$,
where $m_{l}$ is the lepton mass. For the configuration given above,
the stochastic scale $L_{st}$ has different behavior depending on $T$.
At lower temperatures we have
\begin{equation}
\label{e31}
L_{st}\simeq {\left [\frac{\pi^{3/2}\, M_{H}^{2}\, e^{2\,m_{l}/T}}
{ 48\,\alpha(N)\, m^{11/2}_{l}\,T^{3/2}\, \left
(1+\frac{15}{16}\frac{T}{m_{l}}\right )}\right ]}^{\frac{1}{5}},
\end{equation}
where the condition $2\,n\,\beta\,m_{l} >1$  is taken into account for any integer $n$;
$M_{H}$ is the mass of the Higgs-boson.

On the other hand, at higher temperatures, one has
\begin{equation}
 \label{e32}
L_{st}\simeq {\left [\frac{\pi^{2}\, M_{H} ^{2}}{48\,\zeta (3)\, \alpha(N)\,
T^{3}\, m^{4}_{l}}\right ]}^{\frac{1}{5}},\,\,\,\,\,
\zeta (3) = \sum_{n=1}^{\infty} n^{-3} = 1.202.
\end{equation}

In case of two real photons  one can use the longitudinal stochastic scale $L^{long}_{st} (m_{l} = m_{T}/2)$, with the average transverse mass
$  m_{T} = 0.5 (\sqrt {p_{T_{1}}^{2}} + \sqrt {p_{T_{2}}^{2}})$ in
the frame of, e.g., the Longitudinal Center of Mass System (LCMS) [7].

It turns out that the scale $L_{st}$ defines the range of stochastic forces.
This effect is given by $\alpha (N)$-coherence degree which can be estimated from
the experiment through the function $C_{2}(Q)$ as $Q$ close to zero, $C_{2}(0)$
\begin{equation}
 \label{e33}
\alpha (N) = \frac{1 + \gamma^{1/2}(n) -\tilde C_{2}(0) + \gamma^{1/4}(n)
\sqrt {\tilde C_{2}(0) [\gamma^{1/2}(n) -2 ] + 2}}{\tilde C_{2}(0)-1},
\end{equation}
where $\gamma (n)$ is defined in (\ref{e27}) and $\tilde C_{2}(0)\equiv C_{2}(0) /\xi (N)$.
The upper limit of $C_{2}(0)$ depends on $\langle N\rangle$ and the quantum thermal features
of BE correlation pattern given by $\gamma (n)$: $C_{2}(0) < \xi (N) [1 - \gamma^{1/2}(n)/2 ]^{-1}$. This upper limit is restricted by the maximal value of 2 in the ideal case as $\langle N\rangle \rightarrow\infty$ and $\gamma (n) =1$.

Note, that the limit $\alpha\rightarrow\infty$ yields for fully coherent sources with large $\langle N\rangle$, while the $\alpha\rightarrow 0$ case stands for fully chaotic (incoherent) sources for small  $\langle N\rangle $.
The increasing of $T$ leads to squeezing of the region of stochastic forces influence, and $L_{st}(T=T_{0})= R$ at some effective temperature $T_{0}$.
The higher temperatures, $T > T_{0}$,
satisfy to more squeezing effect and at the critical temperature
$T_{c}$ the scale $L_{st}(T=T_{c})$ takes its minimal value.
Obviously, $T_{c}\sim O(200~GeV)$ defines the phase transition where
the electroweak symmetry restoration will occur.
Since in the phase where $T>T_{c}$ all the masses tend to zero and $\alpha\rightarrow 0$, one should expect the sharp expansion of the region with $L_{st}(T>T_{c})\rightarrow \infty$.

Using the relation between $L_{l}$ and $m_{T}$ obtained from the Heisenberg uncertainty relation $L_{l}(m_{T}) = c\sqrt{h\,L_{0}/2\,\pi\,m_{T}}$ [8] at $T = T_{0}$, one can estimate $R_{0} = L_{0}$ within the formula:
\begin{equation}
\label{e34}
R_{0}\simeq \frac{2\,\pi\, m_{T}}{c^{2}\,h}{\left [\frac{\pi^{2}\,M_{H}^{2}}
{3\,\zeta (3)\, \alpha(N)\,T_{0}^{3}\, m_{T}^{4}}\right ]}^{\frac{2}{5}},
\end{equation}
where $c$ and $h$ are the speed of light in vacuum and
the Planck constant, respectively. The expression (\ref{e34}) relates
the two-photon emission time $R_{0}$ with the Higgs-boson mass $M_{H}$, the transverse mass
of two photons $m_{T}$ and the temperature of two photons emission source $T_{0}$.
We obtain $R_{0}\sim 10^{-24} sec$  for $M_{H} = 120$ GeV with $m_{T} = 1$ GeV (two muon pairs in the final state) in the wide interval $T_{0} = 10 - 100$ GeV. 
In Fig.1 we display the Higgs-boson masses for various values of $T_{0}$ and $\alpha$
with $R = 1~fm$ and $m_{l} = m_{\mu} = 105$ MeV.
As an illustration we present here an estimation of $M_{H} = 138$ GeV assuming
$T_{0} = 20$ GeV,  $\alpha = 1$ ($\lambda\simeq 0.8$).

\begin{figure}[h!]
\renewcommand{\figurename}{Fig.}
  \centering
    \includegraphics[width=\textwidth, height = 85mm]{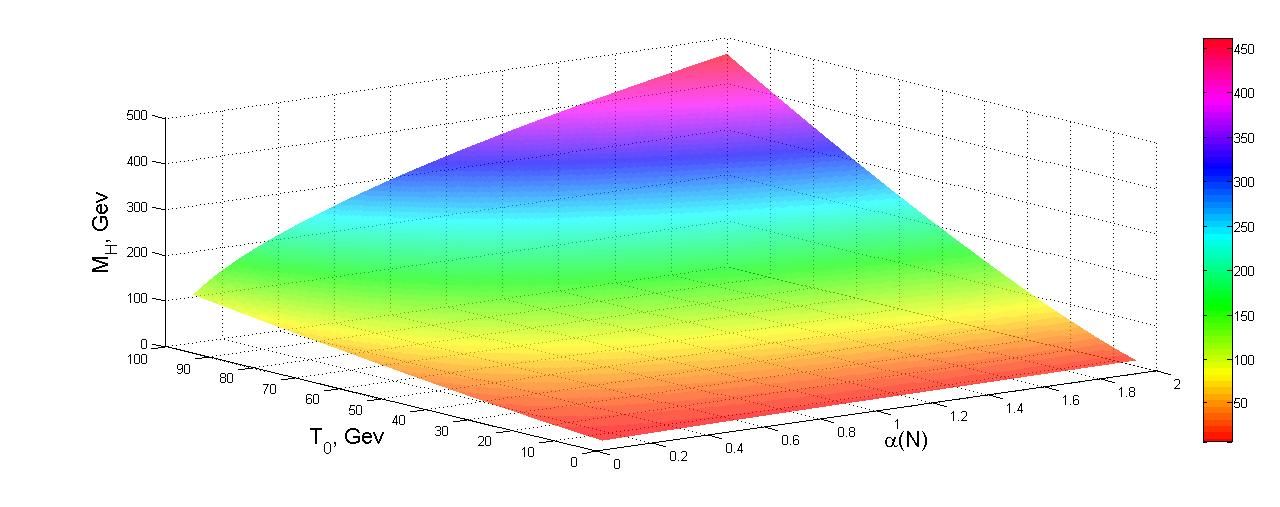}
    \caption{{ \it Higgs-boson masses as a function of $T_{0}$ and $\alpha (N)$
     for $R = 1 fm$. We have set $m_{l} = m_{\mu} = 105~MeV$. }}
  \label{fig:secondgraph}
\end{figure}

The qualitative relation between $R$ and $L_{st}$ above mentioned is the only one we
can emphasize in order to explain the mass and the temperature dependencies of the source size.
The dependence of the stochastic measure of chaoticity $\alpha$ on the minimum scale cut $L_{st}$
and $Q_{rms}$ can be used to define the fit region for different $p_{T}$. Such a minimum
cut on $L_{st}$ introduces a lower cutoff on $R$.

There are  a number of effects which may give rise to $Q\simeq 0$ correlations and, thus,
mimic the two-photons BEC coming from Higgs decays. These include $i)$ correlations of light
hadrons missidentified as photons, $ii)$ radiative decays of resonances in both pseudoscalar and vector sectors, $iii)$ collective flow, etc.

Apparatus or analysis effects which may result in the sensitivity of external random forces
influence, may be investigated by studying the dependence of the correlation functions on
the stochastic scale $L_{st}$. This effect is expected to contribute strongly at small $L_{st}$
(or large $\alpha$ and $T$).

We find that the stochastic scale $L_{st}$ decreases with increasing $T$ slowly at low
temperatures, and it decreases rather abruptly when the critical temperature is approached.
Actually, the experimental measuring of $R$  can provide the precise estimation of $M_{H}$ at
an effective temperature $T_{0}$.
The parameter $\alpha$ can be extracted from the experiment  using $C_{2}(Q)$ with $Q$ being close to zero (\ref{e33}).

Our model is consistent with the idea of the unification of weak and electromagnetic
interactions at $T > T_{c}$, predicted by Kirzhnitz and  Linde in 1972 [9].
In addition, there is the analogy with the asymptotic free theory:
the properties of environment (medium) are the same as those  composed of free particles
in the infinite volume (Universe).

 {\it To summarize:} we faced to the two-photon correlation function  $C_{2}(Q,\beta)$  in which the contribution of $N$, $T$, $\alpha$ are presented. This differs from the methods based on, e.g., the Goldhaber-like approach (\ref{e25}), the trigonometrical parametrization [10], etc.
In particular, by studying the BE correlations of two photons at finite temperature  we can predict the mass of the SM Higgs-boson. 
There is the possibility to determine the precision with which the source size
parameter and the strength chaoticity parameter(s) can be measured at hadron colliders.
Such investigations provide also an opportunity for probing the temperature of the particle production source and the details of in-medium distortion.

I thank I. Gorbunov for help in numerical calculations.
{
\begin{center}
REFERENCES

\end{center}

1. R.M. Weiner, Phys. Rep. 327 (2000) 249.

2. J.F. Gunion, H. E. Haber, G. Kane, S. Dawson, The Higgs Hunter's Guide (Addison Wesley, 1990).

 3. G.A. Kozlov, Phys. Nucl. Part. Lett. 6 (2009) 162; 177.

 4. G.A. Kozlov, Phys. Rev. C58 (1998) 1188;  J. Math.  Phys. 42 (2001) 4749;  New J. of Physics 4 (2002) 23; G.A. Kozlov, O.V. Utyuzh and G. Wilk, Phys. Rev. C68 (2003) 024901;  G.A. Kozlov, J. Elem. Part. Phys. Atom. Nucl. 36 (2005) 108;  G.A. Kozlov, O.V. Utyuzh, G. Wilk, W. Wlodarczyk, Phys. of Atomic Nucl.  71 (2008) 1502.

 5. G. Goldhaber et al., Phys. Rev. Lett. 3 (1959) 181;  G. Goldhaber et al., Phys. Rev. 120 (1960) 300.

6.  V. Khachatryan et al. (CMS Collaboration), Phys. Rev. Lett. 105 (2010) 032001.


7. T. Csorgo, S. Pratt, Proc. Workshop on Rel. Heavy Ion Phys. (Eds. T. Csorgo et al.) 1991 75.

8. G.Alexander, Phys. Lett. B506 (2001) 45; G. Alexander and E.Reinhertz-Aronis, arXiv:0910.0138 [hep-ph].

9. D.A. Kirzhnits, JETP Lett. 15 (1972) 529;  D.A. Kirzhnits, A.D. Linde, Phys. Lett. 42B (1972), 471.

10. T. Csorgo, W. Kittel, W.J. Metzger, T. Novak, Phys. Lett. B663 (2008) 214.

\end{document}